# Fully Automated Noncoplanar Radiation Therapy Treatment Planning

**Charles Huang [1], Yong Yang [2] and Lei Xing [2]**

[1] Department of Bioengineering, Stanford University, Stanford, USA
[2] Department of Radiation Oncology, Stanford University, Stanford, USA

E-mail: xxx@xxx.xx

## Abstract

Noncoplanar radiation therapy treatment planning has the potential to improve dosimetric quality as compared to traditional coplanar techniques. Likewise, automated treatment planning algorithms can reduce a planner's active treatment planning time and remove inter-planner variability. To address the limitations of traditional treatment planning, we have been developing a suite of algorithms called station parameter optimized radiation therapy (SPORT). Within the SPORT suite of algorithms, we propose a method called NC-POPS to produce noncoplanar (NC) plans using the fully automated Pareto Optimal Projection Search (POPS) algorithm. Our NC-POPS algorithm extends the original POPS algorithm to the noncoplanar setting with potential applications to both IMRT and VMAT. The proposed algorithm consists of two main parts: 1) noncoplanar beam angle optimization (BAO) and 2) fully automated inverse planning using the POPS algorithm. We evaluate the performance of NC-POPS by comparing between various noncoplanar and coplanar configurations. To evaluate plan quality, we compute the homogeneity index (HI), conformity index (CI), and dose-volume histogram (DVH) statistics for various organs-at-risk (OARs). As compared to the evaluated coplanar baseline methods, the proposed NC-POPS method achieves significantly better OAR sparing, comparable or better dose conformity, and similar dose homogeneity. Our proposed NC-POPS algorithm provides a modular approach for fully automated treatment planning of noncoplanar IMRT cases with the potential to substantially improve treatment planning workflow and plan quality.

## 1. Introduction

External beam radiation therapy (EBRT) involves the delivery of ionizing radiation to treat diseased tissue while sparing nearby organs-at-risk (OARs) (Galvin et al 2004, Timmerman and Xing 2012). In the case of intensity-modulated radiation therapy (IMRT), planners determine the appropriate plan configuration (i.e. radiation modality, incident beam directions, isocenter location, etc.) and perform inverse planning of individual beam fluences to satisfy various objectives and constraints (Fu et al 2019, Wieser et al 2017). Volumetric modulated arc therapy (VMAT), which allows for continuous rotational delivery of intensity-modulated fields (Otto 2008, Yu 1995, Unkelbach et al 2015, Crooks et al 2003), uses many of these same principles and has been shown to improve delivery efficiency while maintaining or improving dosimetric quality (Rao et al 2010, Verbakel et al 2009, Wolff et al 2009, QUAN et al 2012, Rosenthal et al 2010, Hardcastle et al 2011).

Clinical IMRT and VMAT have traditionally used equispaced coplanar angles and trajectories, where plans are created manually by experienced planners. Traditional techniques assume that incident beam directions are equispaced and coplanar, though plan quality can potentially be improved by adopting more intelligent angle sampling schemes and noncoplanar directions (Li and Xing 2011, 2013, Smyth *et al* 2019). Similarly, manual treatment planning can be time-consuming and labor intensive, and developing automated methods can substantially improve clinical workflow (Hussein *et al* 2018, Huang *et al* 2021). To address the limitations traditional techniques, we have been developing a suite of algorithms called station parameter optimized radiation therapy (SPORT). Within the SPORT suite of algorithms, various algorithms have been developed to address the limitations of equispaced IMRT and VMAT (i.e. segmentally boosted VMAT, DASSIM, etc.) (Li and Xing 2011, 2013, Xing and Li 2014). Moreover, we previously developed an algorithm for fully automated Pareto optimal and clinically acceptable treatment planning in the coplanar setting (Huang *et al* 2021). In this current work, we propose the noncoplanar Pareto optimal projection search (NC-POPS) algorithm for fully automated noncoplanar treatment planning.

### 1.1 Beam Angle Optimization

Beam angle optimization (BAO), critically important to noncoplanar treatment planning, has traditionally not been a part of the coplanar treatment planning process. However, previous studies on noncoplanar IMRT and VMAT have revealed the potential for significant dosimetric improvements when comparing noncoplanar plans to their coplanar counterparts. In this part, we provide a brief overview of BAO approaches and direct interested readers to a more in-depth review for further reading (Smyth *et al* 2019).

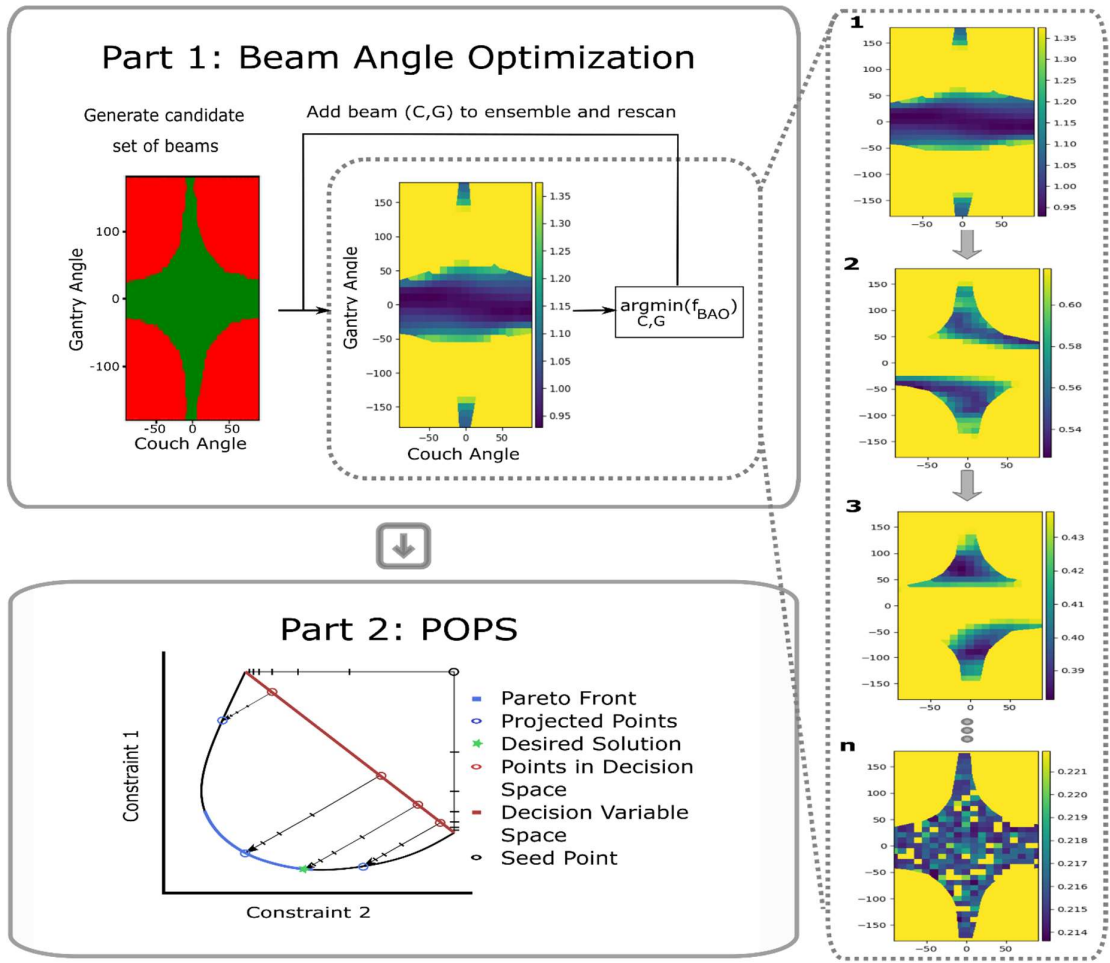

**Figure 1.** Visualizes the overall workflow for the NC-POPS method. In part 1, we perform BAO using an iterative approach. At each iteration η, we exhaustively search for the beam angle which minimizes $F_{BAO}$ and add it to the beam ensemble. Upon completion of part 1, we have an ensemble of $n$ promising beam angles that can then be used to perform automated inverse planning using the POPS algorithm.

$$EUD_s = \left(\frac{1}{N_s}\sum_{i \in s} d_i^{a_s}\right)^{1/a_s} \tag{1}$$

$$F_{BAO} = EUD_{ptv} + \sum_{s \in OARs} p_s\, EUD_s \tag{2}$$

$$\min_x \quad F_{BAO}$$

$$s.t. \quad\quad x \geq 0$$
$$\vec{d} = D\vec{x} \tag{3}$$

**Table 1.** List of parameters used in the beam angle optimization process.

|  | PTV | Rectum | Bladder | FH R | FH L | Body |
|---|---|---|---|---|---|---|
| Penalty ($p_s$) | 1 | 0.2 | 0.2 | 0.03 | 0.03 | 1 |
| Overlap Priority | 1 | 2 | 2 | 3 | 3 | 3 |
| EUD a | -20 | 1 | 1 | 1 | 1 | 5 |
| Ref. Dose (Gy) | 74 | 0 | 0 | 0 | 0 | 0 |

BAO approaches are typically categorized into two different classes. The first class of approaches does not incorporate fluence map optimization (FMO) into the beam angle selection (BAS) process. Instead, these approaches typically select beam angles using heuristics based on geometric or dosimetric information (Bangert and Oelfke 2010, Smyth *et al* 2013, MacDonald and Thomas 2015, Yang *et al* 2011, Fix *et al* 2018). In contrast, the second class of approaches does incorporate FMO into the BAS process. For this class of methods, the BAS problem is typically formulated as a combinatorial optimization (CO) problem, where an ensemble of promising beam angles is selected from the feasible set of candidate beam angles (i.e. beam angles where there are no couch-gantry collisions). A comparison of several strategies for solving the CO problem was done previously by Bangert et al. (Bangert *et al* 2013), with many alternative strategies being proposed as well (Bangert *et al* 2013, Stein *et al* 1997, Pugachev *et al* 2001, Papp *et al* 2015, Hou *et al* 2003, Wu *et al* 2000, Li *et al* 2004, Wang *et al* 2003, Li *et al* 2005, Woudstra and Storchi 2000, Pugachev and Xing 2002).

### 1.2 Pareto Optimal Projection Search

The workflow for treatment planning, in general, has been a manual, iterative process performed by human planners. This iterative planning process is not only time-consuming and labor intensive, but the resulting plan quality varies with planner skill and experience (Clark *et al* 2008, Shen *et al* 2020, Roach *et al* 2019, Nelms *et al* 2012, Berry *et al* 2016, Batumalai *et al* 2013, Moore *et al* 2012, Xing *et al* 1999). For instance, in the case of prostate treatment planning, our clinical protocol usually places high priority on sparing organs like the rectum and bladder, with less priority placed on sparing organs like the femoral heads. However, individual planner preferences may vary, which can lead to substantial variations in plan quality. Addressing these issues by reducing active treatment planning time and inter-planner variability has tremendous potential for improving clinical workflow and quality of care.

Previously, we developed the Pareto Optimal Projection Search (POPS) algorithm (Huang *et al* 2021) to automate

treatment planning workflow by producing plans that are both Pareto optimal and clinically acceptable. The original POPS algorithm was applied to the coplanar setting. In this paper, we provide substantial extensions of the algorithm to the noncoplanar setting by adapting an iterative approach (Bangert *et al* 2013, Papp *et al* 2015) to the CO problem.

## 2. Methods

Here, we propose the noncoplanar Pareto optimal projection search (NC-POPS) algorithm, which consists of two main steps: 1) BAO and 2) fully automated inverse planning using POPS. In this work, we focus on applying the proposed method to the noncoplanar IMRT setting. The proposed method, however, can also be applied to the noncoplanar VMAT setting (Papp *et al* 2015).

First, BAO is performed by solving the CO problem using an iterative approach with limited number of FMO iterations (Papp *et al* 2015). This iterative approach produces an ensemble of promising beam directions, which we call beam control points, to be used for noncoplanar IMRT planning. For noncoplanar VMAT, these beam control points are then converted to dynamic trajectories by solving a shortest path problem between beam control points, as described by Papp et al. (Papp *et al* 2015).

Second, given the set of beam control points or dynamic trajectories, fully automated inverse planning can be performed by applying the POPS algorithm. POPS searches the feasibility boundary, which contains the Pareto front, for the most desirable plans as determined by a given scoring function. In this work, we use a normalized equivalent uniform dose (nEUD), which normalizes the EUD of each OAR by the maximum dose.

### 2.1 BAO Using an Iterative Approach

The BAO process can be formulated as a combinatorial optimization problem and is visualized in part 1 of Figure 1. In this formulation, we attempt to create an ensemble of promising beam directions by iteratively adding beam angles (i.e. couch and gantry angle pairs) from the candidate set of feasible beams. We begin by determining which couch and gantry angle pairs are feasible and which lead to undesirable collisions. Next, an iterative approach, or "lookahead strategy," is adopted to select promising beam angles (Bangert *et al* 2013, Papp *et al* 2015). In each iteration η of the approach, we can compute the utility of each feasible beam angle as defined by an objective function. This is done by performing fluence map optimization (FMO) for each beam angle. For our work, we use the objective function $F_{BAO}$, which contains equivalent uniform dose objectives for the target and various OARs. Definitions for EUD and $F_{BAO}$ are shown in Equations 1 and 2. $p_i$ and $d_i$ refer to the penalty and dose, respectively, at voxel $i$ in a particular structure $s$. $N_s$ refers to the total number of voxels in the structure $s$, and $a_s$ is a structure-specific EUD parameter. Other popular choices for this objective function include quadratic dose objective terms for the target, OARs, and distance-based dose conformity.

In the first iteration ($\eta = 1$) of the approach, we exhaustively search for the beam angle which provides the minimum objective function value and add the selected beam to the ensemble (denoted as $BE_\eta$). In the next iteration $\eta + 1$, we compute the objective function values for combinations of each beam in the feasible set with the existing ensemble $BE_\eta$.

We again search for the beam angle with the minimum objective function value and add it to the ensemble $BE_{\eta+1}$. This iterative approach, similar to previous implementations (Bangert *et al* 2013, Papp *et al* 2015), is repeated until the beam ensemble contains the desired number of beams.

In part 1 of Figure 1, we provide a visualization of the iterative approach for BAO. Each iteration of the approach involves computing objective function values for the candidate beam angle combinations and adding the minimum beam angle combination to the beam ensemble. In Table 1, we list the parameters relevant to the BAO process. To be consistent with our clinical protocol, higher emphasis (larger $p_s$ values) is placed on sparing the rectum and bladder, and less emphasis is placed on sparing the femoral heads. Additionally, higher emphasis is placed on the body objective

(EUD with $\alpha = 5$), which acts as an indirect objective for dose conformity.

## 2.2 POPS for Fully Automated Inverse Planning

The POPS algorithm provides a general framework for fully automated inverse planning (Huang *et al* 2021) and is visualized for a hypothetical 2D case in part 2 of Figure 1. Its main purposes are to produce plans that are Pareto optimal and clinically acceptable, while requiring no active planning time during the inverse planning process. Pareto optimal plans, or efficient non-dominated plans, are critically important to high quality treatment planning. Traditionally, multicriteria optimization approaches have focused on generating a diverse set of Pareto optimal plans (i.e. the Pareto front) and delegated the selection which plans are clinically acceptable to a human planner. The POPS algorithm automates this selection process by searching the feasibility boundary (which contains the Pareto front) for the most desirable plans, as defined by a scoring function. For a detailed description of the POPS algorithm, please see the original POPS paper (Huang *et al* 2021). For this current study, we additionally utilize the open-source treatment planning software package called MatRad (Wieser *et al* 2017).

$$nEUD_s = \frac{EUD_s}{D_{max}} \tag{4}$$

$$F_{POPS} = \begin{cases} 1, & if\ infeasible \\ \frac{\sum_{s \in S} w_s \cdot nEUD_s}{\sum_{s \in S} w_s}, & otherwise \end{cases} \tag{5}$$

In this current work, we use a normalized equivalent uniform dose (nEUD) metric to score plans for the POPS algorithm, as shown in Equations 4 and 5. Here, $EUD_s$ refers to the equivalent uniform dose of the structure $s$, $D_{max}$ refers to the maximum dose ($1.05D_p$, where $D_P$ refers to the prescription dose), and $F_{POPS}$ is the scoring function used by the POPS algorithm to search the feasibility boundary. Following the POPS algorithm (Huang *et al* 2021), we formulate the inverse planning problem as a feasibility search,

**Table 2.** List of parameters used during POPS automated inverse planning. Prescription dose $D_p = 74\ Gy$. Plan scoring ($F_{POPS}$) is performed using Equations 4 and 5, and the feasibility search is formulated in Equation 6.

|  | PTV | Rectum | Bladder | FH R | FH L | Body |
|---|---|---|---|---|---|---|
| Score Weight ($w_s$) | N/A | 1 | 1 | 0.15 | 0.15 | 5 |
| Overlap Priority | 1 | 2 | 2 | 3 | 3 | 3 |
| EUD $\alpha$ | N/A | 1 | 1 | 1 | 1 | 5 |

$$\begin{aligned}
\min_x \quad & \frac{1}{N_{ptv}} \sum_{i \in ptv} (d_i - D_p)^2 \\
s.t. \quad & x \geq 0 \\
& \vec{d} = \boldsymbol{D}\vec{x} \\
& EUD_{rectum} \leq c_1 \\
& EUD_{bladder} \leq c_2 \\
& EUD_{FH\ R} \leq c_3 \\
& EUD_{FH\ L} \leq c_4 \\
& EUD_{body} \leq c_5 \\
& D_{ptv}(95\%) \geq D_p \\
& D_{ptv}(min) = 0.975 D_p \\
& D_{ptv}(max) = 1.05 D_p
\end{aligned} \tag{6}$$

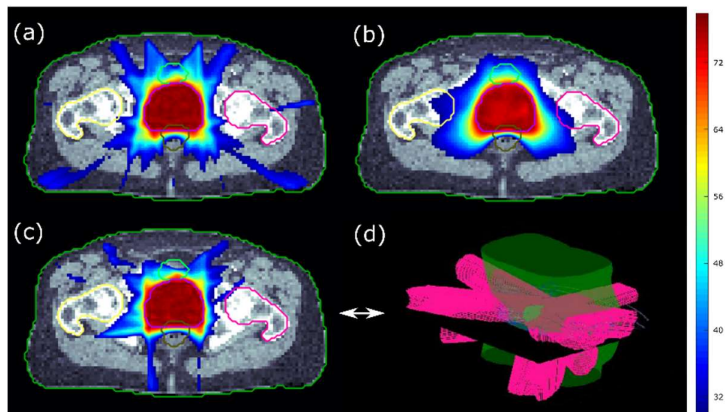

**Figure 2.** Example dose distributions are visualized for the (a) POPS coplanar IMRT plan, (b) Manual coplanar VMAT plan, and (c) NC-POPS IMRT plan. Part (d) provides a visualization of the 3D rendered noncoplanar beam directions.

as shown in Equation 6. For all plans generated using POPS, we use a prescription dose $D_p$ of 74 Gy. Table 2 lists the relevant parameters used during the POPS algorithm. For a detailed description of the POPS algorithm and its implementation, we refer interested readers to the original POPS paper (Huang *et al* 2021).

## 3. Results

### 3.1 Experimental Setup and Evaluation

We study the performance of our approach on a dataset of 21 prostate cases (Stanford IRB protocol #41335). Two baseline methods are included for comparison. For the first baseline method, clinical VMAT plans with two full coplanar arcs were manually planned by experienced planners. For the second baseline method, 9-beam coplanar IMRT plans were generated automatically using the POPS algorithm. These baseline methods are compared against plans using our proposed method, NC-POPS, configured for 9-beam noncoplanar IMRT.

Treatment plans were evaluated based on dose conformity, dose homogeneity, and OAR sparing (Yang *et al* 2020, Pyakuryal *et al* 2010, Ventura *et al* 2016, Cao *et al* 2019, Kataria *et al* 2012, van't Riet *et al* 1997, Semenenko *et al* 2008, Paddick 2000). We adopt the following definitions (Equations 7 and 8) of the conformity index (CI) (Paddick 2000) and homogeneity index (HI) (Semenenko *et al* 2008):

$$CI = \frac{\left(TV_{95D_p}\right)^2}{TV \times V_{95D_p}} \qquad (7)$$

$$HI = \frac{D_5 - D_{95}}{D_p} \qquad (8)$$

In order to assess OAR sparing, we compute the mean dose and dose-volume histogram (DVH) values at the following percentiles: D(20%), D(40%), D(60%), D(80%), and D(98%). Here, lower dosage values are desirable as they imply better OAR sparing.

### 3.2 Qualitative Comparison

We first conduct a qualitative comparison between the two types of coplanar baseline plans and the NC-POPS plans. A visualization of the differences in dose distributions is provided in Figure 2. Comparing the dose distribution for the proposed NC-POPS method to either of the coplanar baselines, we can visually appreciate a substantial improvement in OAR sparing with comparable conformity and homogeneity. In Figure 2d, we visualize a rendering of the beam ensemble created using BAO. The ensemble of promising beam angles is highly noncoplanar, with fewer OAR voxels in the paths of the selected beams than in the coplanar setting.

Figure 3 provides DVH comparisons between NC-POPS and the coplanar baselines (see Figure 3a and Figure 3b) for six example patients (numbered 1-6). In Figure 3a, we compare the performance of NC-POPS plans to that of the manual coplanar VMAT plans. In each of the six patients, OAR sparing for all organs is significantly improved by the proposed method. In Figure 3b, we compare the performance of NC-POPS plans to the performance of coplanar POPS plans for the same six patients shown in Figure 3a. OAR sparing of the proposed NC-POPS method is significantly better for all structures except the body. PTV performance was comparable between NC-POPS and the coplanar baselines for all example patients.

### 3.3 Quantitative Comparison

In Table 3, the proposed NC-POPS method is compared to the two coplanar baseline methods. Plan quality is assessed using the conformity index, the homogeneity index, and various OAR sparing statistics (mean dose, D(20%), D(40%), D(60%), D(80%), and D(98%)). Here values like $D(\cdot\%)$ represent metrics for the DVH and can be interpreted as $(\cdot\%)$ of the structure receives at least $D(\cdot\%)$ dose. Differences between the evaluated methods were quantified using the Wilcoxon signed-rank test ($p < 0.05$).

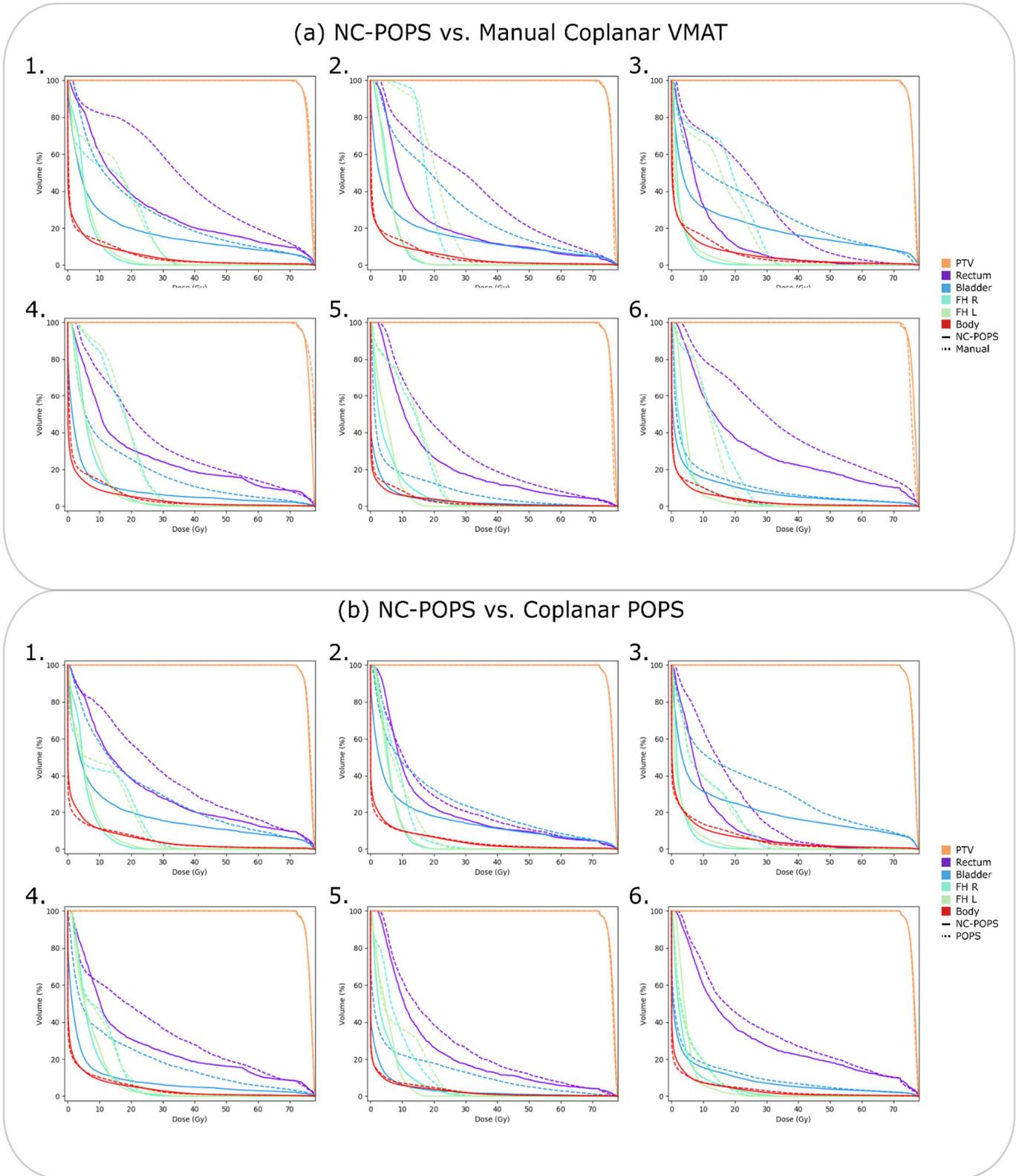

**Figure 3.** DVHs are visualized for six example patients (numbers 1-6). In part (a), we compare our proposed NC-POPS IMRT plans (solid lines) to manual coplanar VMAT plans (dashed lines). In part (b), we compare our proposed NC-POPS IMRT plans (solid lines) to coplanar POPS IMRT plans (dashed lines). While achieving comparable PTV performance, the proposed NC-POPS IMRT plans provide substantially better OAR sparing than either the manual coplanar VMAT plans or the coplanar POPS IMRT plans.

**Table 3.** DVH values at five percentiles (D(20%), D(40%), D(60%), D(80%), D(98%)) were used to compare OAR sparing. Conformity and homogeneity are additionally assessed. The best values are bolded for readability.

| | Conformity Index (CI) | Homogeneity Index (HI) | OAR | Mean Dose ($\mu$) | D(20%) (Gy) | D(40%) (Gy) | D(60%) (Gy) | D(80%) (Gy) | D(98%) (Gy) |
|---|---|---|---|---|---|---|---|---|---|
| Manual Coplanar VMAT | 0.86 (0.03) | 0.05 (0.01) | Rectum | 30.2 (5.6) | 51.2 (10.6) | 31.2 (7.6) | 20.0 (6.9) | 9.4 (5.0) | 3.0 (1.0) |
| | | | Bladder | 19.2 (9.5) | 34.8 (20.7) | 14.5 (11.4) | 6.6 (6.8) | 3.7 (4.9) | 1.9 (2.5) |
| | | | FH R | 16.4 (3.4) | 23.6 (4.7) | 19.3 (4.2) | 14.4 (4.0) | 7.1 (4.2) | 2.2 (2.9) |
| | | | FH L | 15.0 (3.1) | 21.9 (4.6) | 17.0 (3.3) | 13.0 (3.2) | 6.8 (4.2) | 2.1 (2.8) |
| | | | Body | 3.9 (0.8) | 3.3 (1.6) | 0.5 (0.2) | 0.2 (0.1) | 0.1 (0.0) | 0* |
| Coplanar POPS 9-Beam IMRT | 0.90 (0.03) | 0.05 (0.00) | Rectum | 26.0 (6.3) | 46.2 (11.7) | 25.7 (8.7) | 14.0 (5.9) | 6.4 (3.9) | 2.4 (1.9) |
| | | | Bladder | 19.0 (8.8) | 36.5 (20.0) | 13.7 (11.8) | 5.8 (6.6) | 3.0 (4.3) | 1.3 (2.4) |
| | | | FH R | 8.6 (2.2) | 15.6 (3.9) | 8.7 (3.7) | 4.2 (2.1) | 2.1 (1.0) | **0.6 (0.5)** |
| | | | FH L | 8.6 (2.3) | 16.1 (4.2) | 8.2 (3.9) | 3.8 (1.5) | 1.9 (0.9) | **0.5 (0.5)** |
| | | | Body | 3.7 (0.8) | **2.4 (1.6)** | **0.2 (0.3)** | 0.1 (0.0) | 0* | 0* |
| NC-POPS 9-Beam IMRT | 0.90 (0.04) | 0.05 (0.00) | Rectum | **21.9 (6.3)** | **38.1 (14.8)** | **18.1 (7.6)** | **10.1 (3.6)** | **5.7 (2.0)** | **2.2 (1.2)** |
| | | | Bladder | **14.3 (8.7)** | **25.1 (22.2)** | **8.5 (9.4)** | **3.5 (4.1)** | **1.7 (2.3)** | **0.6 (1.3)** |
| | | | FH R | **5.7 (1.5)** | **9.1 (2.9)** | **5.1 (1.6)** | **3.0 (1.2)** | **1.7 (0.8)** | 0.7 (0.4) |
| | | | FH L | **5.7 (1.4)** | **8.8 (2.5)** | **5.5 (1.6)** | **3.3 (1.2)** | 1.9 (1.0) | 0.8 (0.6) |
| | | | Body | 3.7 (0.8) | 3.3 (1.6) | 0.4 (0.4) | 0.1 (0.0) | 0* | 0* |

| | | Conformity Index (CI) | Homogeneity Index (HI) | Rectum ($\mu$) | Bladder ($\mu$) | FH R ($\mu$) | FH L ($\mu$) | Body ($\mu$) |
|---|---|---|---|---|---|---|---|---|
| NC-POPS vs. Manual Coplanar VMAT | Wilcoxon Signed-rank Test (p-value) | **0.00313** | 0.12717 | **0.00006** | **0.00014** | **0.00006** | **0.00006** | **0.01295** |
| NC-POPS vs. Coplanar POPS | Wilcoxon Signed-rank Test (p-value) | 0.83479 | 0.54617 | **0.00006** | **0.00028** | **0.00006** | **0.00006** | 0.35701 |

*Values were vanishingly small

Dose conformity values are summarized in Table 3, with CI values approaching 1 in an ideal case. The proposed NC-POPS method achieves significantly better conformity than the manual coplanar baseline and comparable conformity to the POPS coplanar baseline (p-values of 0.00313 and 0.83479, respectively). Dose homogeneity values are additionally compared in Table 3, where HI approaches zero for an ideal case. Dose homogeneity values for the evaluated methods were comparable.

Finally, OAR sparing is assessed using the mean dose and various DVH metrics. The proposed NC-POPS method achieves the best OAR sparing among the evaluated methods with significantly lower mean doses to the rectum, bladder, and femoral heads than either of the two coplanar baseline methods. Moreover, the proposed method achieves the lowest values in most of the OAR DVH metrics that were evaluated. Overall, the proposed NC-POPS method achieves better OAR sparing, comparable or better dose conformity, and comparable dose homogeneity when compared to the two coplanar baseline methods.

## 4. Discussion

This study introduced the NC-POPS method for fully automated noncoplanar treatment planning. NC-POPS provides substantial extensions of the original POPS algorithm to the noncoplanar setting by incorporating BAO using an iterative algorithm. The performance of the NC-POPS method for IMRT was compared to two coplanar baseline methods (coplanar POPS IMRT and manual coplanar VMAT). Here, we evaluated the performance of our proposed method on a dataset of prostate cases, but we anticipate that the proposed method can be readily applied to the treatment planning of other sites in the body, with slight modifications to Equations 3 and 6.

Here, we implemented an iterative approach to BAO, though many alternative approaches to the BAO problem also exist (e.g. simulated annealing, genetic algorithms, cross-entropy algorithms, etc.). While comprehensive comparisons of various BAO approaches are scarce, previous works like Bangert et al. (Bangert *et al* 2013) conclude that the performance of various BAO approaches is similar. Bangert et al. argues that this comparable performance is likely due to convergence of these various methods to solutions that have similar BAO objective function values. In this study, we adopt the iterative approach to BAO, or "lookahead strategy," developed previously due to its simplicity and improved computational efficiency (Bangert *et al* 2013, Papp *et al* 2015).

To perform fully automated treatment planning, the proposed NC-POPS method incorporates the POPS algorithm to search the feasibility boundary, which contains the Pareto front, for the most desirable plans. Plan scoring is performed using the nEUD metric, where high priority is placed on sparing organs like the rectum and bladder, and low priority is placed on sparing organs like the femoral heads. These specific organ priorities are adopted to mimic our clinical protocol.

The proposed NC-POPS method achieves significantly better OAR sparing, significantly better dose conformity than the manual coplanar baseline, and comparable homogeneities to both coplanar baseline methods. Moreover, the two main components of NC-POPS, namely BAO and POPS, are both modular and can be incorporated into existing clinical workflows. The framework can also be readily upgraded to incorporate new developments to either BAO or automated treatment planning.

Moving forward, we hope to apply the current framework to the noncoplanar VMAT setting, as well as to treatment planning of other sites in the body (i.e. head and neck, abdomen, etc.). The BAO component of our framework has previously been applied to manual noncoplanar IMRT and manual noncoplanar VMAT. In the case of noncoplanar VMAT, the ensemble of promising beam directions produced by BAO can be converted to a dynamic trajectory by solving a traveling salesman problem (Papp *et al* 2015). Similarly, applying POPS to automated VMAT treatment planning can be done by applying additional steps after FMO, such as leaf sequencing and direct aperture optimization (Otto 2008, Christiansen *et al* 2018, Papp and Unkelbach 2014).

## 5. Conclusion

In this work, we proposed the NC-POPS framework for fully automated noncoplanar treatment planning. The first component of the framework, BAO, is implemented using an iterative algorithm to produce an ensemble of promising noncoplanar beam directions. These beam directions are then used in the second component of the framework, POPS, to perform fully automated inverse planning. Our experimental results demonstrate that the proposed NC-POPS significantly improves OAR sparing as compared to coplanar methods, while maintaining comparable or better dose conformity and comparable homogeneity. We anticipate that the proposed method will substantially improve clinical workflow and plan quality.

## Acknowledgements

This research was supported by the National Institutes of Health (NIH) under Grants 1R01 CA176553, R01CA227713, and T32EB009653, as well as a faculty research award from Google Inc. The authors would like to thank and acknowledge the MatRad development team for their help and advice.

## References

Bangert M and Oelfke U 2010 Spherical cluster analysis for beam angle optimization in intensity-modulated radiation therapy treatment planning *Phys. Med. Biol.* **55** 6023–37

Bangert M, Ziegenhein P and Oelfke U 2013 Comparison of beam angle selection strategies for intracranial IMRT *Med. Phys.* **40** 011716

Batumalai V, Jameson M G, Forstner D F, Vial P and Holloway L C 2013 How important is dosimetrist experience for intensity modulated radiation therapy? A comparative analysis of a head and neck case *Pract. Radiat. Oncol.* **3** e99–106

Berry S L, Boczkowski A, Ma R, Mechalakos J and Hunt M 2016 Interobserver variability in radiation therapy plan output: Results of a single-institution study *Pract. Radiat. Oncol.* **6** 442–9

Cao T, Dai Z, Ding Z, Li W and Quan H 2019 Analysis of different evaluation indexes for prostate stereotactic body radiation therapy plans: conformity index, homogeneity index and gradient index *Precis. Radiat. Oncol.* **3** 72–9

Christiansen E, Heath E and Xu T 2018 Continuous aperture dose calculation and optimization for volumetric modulated arc therapy *Phys. Med. Biol.* **63** 21NT01

Clark V H, Chen Y, Wilkens J, Alaly J R, Zakaryan K and Deasy J O 2008 IMRT treatment planning for prostate cancer using prioritized prescription optimization and mean-tail-dose functions *Linear Algebra Its Appl.* **428** 1345–64

Crooks S M, Wu X, Takita C, Watzich M and Xing L 2003 Aperture modulated arc therapy *Phys. Med. Biol.* **48** 1333–44

Fix M K, Frei D, Volken W, Terribilini D, Mueller S, Elicin O, Hemmatazad H, Aebersold D M and Manser P 2018 Part 1: Optimization and evaluation of dynamic trajectory radiotherapy *Med. Phys.*

Fu A, Ungun B, Xing L and Boyd S 2019 A convex optimization approach to radiation treatment planning with dose constraints *Optim. Eng.* **20** 277–300

Galvin J M, Ezzell G, Eisbrauch A, Yu C, Butler B, Xiao Y, Rosen I, Rosenman J, Sharpe M, Xing L, Xia P, Lomax T, Low D A and Palta J 2004 Implementing IMRT in clinical practice: a joint document of the American Society for Therapeutic Radiology and Oncology and the American Association of Physicists in Medicine *Int. J. Radiat. Oncol. Biol. Phys.* **58** 1616–34

Hardcastle N, Tomé W A, Foo K, Miller A, Carolan M and Metcalfe P 2011 Comparison of prostate IMRT and VMAT biologically optimised treatment plans *Med. Dosim. Off. J. Am. Assoc. Med. Dosim.* **36** 292–8

Hou Q, Wang J, Chen Y and Galvin J M 2003 Beam orientation optimization for IMRT by a hybrid method of the genetic algorithm and the simulated dynamics *Med. Phys.* **30** 2360–7

Huang C, Yang Y, Panjwani N, Boyd S and Xing L 2021 Pareto Optimal Projection Search (POPS): Automated Radiation Therapy Treatment Planning by Direct Search of the Pareto Surface *IEEE Trans. Biomed. Eng.* 1–1

Hussein M, Heijmen B J M, Verellen D and Nisbet A 2018 Automation in intensity modulated radiotherapy treatment planning—a review of recent innovations *Br. J. Radiol.* **91** 20180270

Kataria T, Sharma K, Subramani V, Karrthick K P and Bisht S S 2012 Homogeneity Index: An objective tool for assessment of conformal radiation treatments *J. Med. Phys. Assoc. Med. Phys. India* **37** 207–13

Li R and Xing L 2013 An adaptive planning strategy for station parameter optimized radiation therapy (SPORT): Segmentally boosted VMAT *Med. Phys.* **40** Online: https://www.ncbi.nlm.nih.gov/pmc/articles/PMC3656955/

Li R and Xing L 2011 Bridging the gap between IMRT and VMAT: Dense angularly sampled and sparse intensity modulated radiation therapy *Med. Phys.* **38** 4912–9

Li Y, Yao D, Yao J and Chen W 2005 A particle swarm optimization algorithm for beam angle selection in intensity-modulated radiotherapy planning *Phys. Med. Biol.* **50** 3491–514

Li Y, Yao J and Yao D 2004 Automatic beam angle selection in IMRT planning using genetic algorithm *Phys. Med. Biol.* **49** 1915–32

MacDonald R L and Thomas C G 2015 Dynamic trajectory-based couch motion for improvement of radiation therapy trajectories in cranial SRT *Med. Phys.* **42** 2317–25

Moore K L, Brame R S, Low D A and Mutic S 2012 Quantitative Metrics for Assessing Plan Quality *Semin. Radiat. Oncol.* **22** 62–9

Nelms B E, Robinson G, Markham J, Velasco K, Boyd S, Narayan S, Wheeler J and Sobczak M L 2012 Variation in external beam treatment plan quality: An inter-institutional study of planners and planning systems *Pract. Radiat. Oncol.* **2** 296–305

Otto K 2008 Volumetric modulated arc therapy: IMRT in a single gantry arc. *Med. Phys.* **35** 310–7

Paddick I 2000 A simple scoring ratio to index the conformity of radiosurgical treatment plans. Technical note *J. Neurosurg.* **93 Suppl 3** 219–22

Papp D, Bortfeld T and Unkelbach J 2015 A modular approach to intensity-modulated arc therapy optimization with noncoplanar trajectories *Phys. Med. Biol.* **60** 5179–98

Papp D and Unkelbach J 2014 Direct leaf trajectory optimization for volumetric modulated arc therapy planning with sliding window delivery *Med. Phys.* **41** 011701

Pugachev A, Li J G, Boyer A L, Hancock S L, Le Q-T, Donaldson S S and Xing L 2001 Role of beam orientation optimization in intensity-modulated radiation therapy *Int. J. Radiat. Oncol. Biol. Phys.* **50** 551–60

Pugachev A and Xing L 2002 Incorporating prior knowledge into beam orientation optimization in IMRT *Int. J. Radiat. Oncol. Biol. Phys.* **54** 1565–74

Pyakuryal A, Myint W K, Gopalakrishnan M, Jang S, Logemann J A and Mittal B B 2010 A computational tool for the efficient analysis of dose-volume histograms for radiation therapy treatment plans *J. Appl. Clin. Med. Phys.* **11** 137–57

QUAN E M, LI X, LI Y, WANG X, KUDCHADKER R J, JOHNSON J L, KUBAN D A, LEE A K and ZHANG X 2012 A comprehensive comparison of IMRT and VMAT plan quality for prostate cancer treatment *Int. J. Radiat. Oncol. Biol. Phys.* **83** 1169–78

Rao M, Yang W, Chen F, Sheng K, Ye J, Mehta V, Shepard D and Cao D 2010 Comparison of Elekta VMAT with helical tomotherapy and fixed field IMRT: plan quality, delivery efficiency and accuracy. *Med. Phys.* **37** 1350–9

van't Riet A, Mak A C, Moerland M A, Elders L H and van der Zee W 1997 A conformation number to quantify the degree of conformality in brachytherapy and external beam irradiation: application to the prostate *Int. J. Radiat. Oncol. Biol. Phys.* **37** 731–6

Roach D, Wortel G, Ochoa C, Jensen H R, Damen E, Vial P, Janssen T and Hansen C R 2019 Adapting automated treatment planning configurations across international centres for prostate radiotherapy *Phys. Imaging Radiat. Oncol.* **10** 7–13

Rosenthal S A, Wu C, Mangat J K, Tunnicliff C J, Chang G C, Dutton S C, Goldsmith B J, Leibenhaut M H, Logsdon M

D and Asche D R 2010 Comparison of Volumetric Modulated Arc Therapy (VMAT) vs. Fixed Field Intensity Modulated Radiation Therapy (IMRT) Techniques for the Treatment of Localized Prostate Cancer *Int. J. Radiat. Oncol. Biol. Phys.* **78** S755

Semenenko V A, Reitz B, Day E, Qi X S, Miften M and Li X A 2008 Evaluation of a commercial biologically based IMRT treatment planning system *Med. Phys.* **35** 5851–60

Shen C, Nguyen D, Chen L, Gonzalez Y, McBeth R, Qin N, Jiang S B and Jia X 2020 Operating a treatment planning system using a deep-reinforcement learning-based virtual treatment planner for prostate cancer intensity-modulated radiation therapy treatment planning *Med. Phys.*

Smyth G, Bamber J C, Evans P M and Bedford J L 2013 Trajectory optimization for dynamic couch rotation during volumetric modulated arc radiotherapy *Phys. Med. Biol.* **58** 8163–77

Smyth G, Evans P M, Bamber J C and Bedford J L 2019 Recent developments in non-coplanar radiotherapy *Br. J. Radiol.* **92** Online: https://www.ncbi.nlm.nih.gov/pmc/articles/PMC6580906/

Stein J, Mohan R, Wang X-H, Bortfeld T, Wu Q, Preiser K, Ling C C and Schlegel W 1997 Number and orientations of beams in intensity-modulated radiation treatments *Med. Phys.* **24** 149–60

Timmerman R D and Xing L 2012 *Image-guided and adaptive radiation therapy* (Lippincott Williams & Wilkins)

Unkelbach J, Bortfeld T, Craft D, Alber M, Bangert M, Bokrantz R, Chen D, Li R, Xing L, Men C, Nill S, Papp D, Romeijn E and Salari E 2015 Optimization approaches to volumetric modulated arc therapy planning *Med. Phys.* **42** 1367–77

Ventura T, Lopes M do C, Ferreira B C and Khouri L 2016 SPIDERplan: A tool to support decision-making in radiation therapy treatment plan assessment *Rep. Pract. Oncol. Radiother.* **21** 508–16

Verbakel W F A R, Cuijpers J P, Hoffmans D, Bieker M, Slotman B J and Senan S 2009 Volumetric intensity-modulated arc therapy vs. conventional IMRT in head-and-neck cancer: a comparative planning and dosimetric study. *Int. J. Radiat. Oncol. Biol. Phys.* **74** 252–9

Wang C, Dai J and Hu Y 2003 Optimization of beam orientations and beam weights for conformal radiotherapy using mixed integer programming *Phys. Med. Biol.* **48** 4065–76

Wieser H-P, Cisternas E, Wahl N, Ulrich S, Stadler A, Mescher H, Müller L-R, Klinge T, Gabrys H, Burigo L, Mairani A, Ecker S, Ackermann B, Ellerbrock M, Parodi K, Jäkel O and Bangert M 2017 Development of the open-source dose calculation and optimization toolkit matRad *Med. Phys.* **44** 2556–68

Wolff D, Stieler F, Welzel G, Lorenz F, Abo-Madyan Y, Mai S, Herskind C, Polednik M, Steil V, Wenz F and Lohr F 2009 Volumetric modulated arc therapy (VMAT) vs. serial tomotherapy, step-and-shoot IMRT and 3D-conformal RT for treatment of prostate cancer. *Radiother. Oncol. J. Eur. Soc. Ther. Radiol. Oncol.* **93** 226–33

Woudstra E and Storchi P R M 2000 Constrained treatment planning using sequential beam selection *Phys. Med. Biol.* **45** 2133–49

Wu X, Zhu Y, Dai J and Wang Z 2000 Selection and determination of beam weights based on genetic algorithms for conformal radiotherapy treatment planning *Phys. Med. Biol.* **45** 2547–58

Xing L, Li J G, Donaldson S, Le Q T and Boyer A L 1999 Optimization of importance factors in inverse planning *Phys. Med. Biol.* **44** 2525–36

Xing L and Li R 2014 Inverse planning in the age of digital LINACs: station parameter optimized radiation therapy (SPORT) *J. Phys. Conf. Ser.* **489** 012065

Yang Y, Shao K, Zhang J, Chen M, Chen Y and Shan G 2020 Automatic Planning for Nasopharyngeal Carcinoma Based on Progressive Optimization in RayStation Treatment Planning System: *Technol. Cancer Res. Treat.* Online: https://journals.sagepub.com/doi/10.1177/1533033820915710

Yang Y, Zhang P, Happersett L, Xiong J, Yang J, Chan M, Beal K, Mageras G and Hunt M 2011 Choreographing Couch and Collimator in Volumetric Modulated Arc Therapy *Int. J. Radiat. Oncol. Biol. Phys.* **80** 1238–47

Yu C X 1995 Intensity-modulated arc therapy with dynamic multileaf collimation: an alternative to tomotherapy *Phys. Med. Biol.* **40** 1435–49